# Estimation of erythrocyte surface area in mammals

Ion Udroiu

*Dipartimento di Scienze, Università degli Studi Roma Tre*

*Viale Marconi 446, 00146 Roma, Italy*

**ABSTRACT**

Measures of erythrocytes volume and surface are helpful in several physiological studies, both for zoologists and veterinarians. Whilst diameter and volume are assessed with ease from observations of blood smears and complete blood count, respectively, thickness and surface area, instead, are much more difficult to be obtained.

The accurate description of the erythrocyte geometry is given by the equation of the oval of Cassini, but the formulas deriving from it are very complex, comprising elliptic integrals. In this article three solids are proposed as models approximating the erythrocyte: sphere, cylinder and a spheroid with concave caps. Volumes and Surface Areas obtained with these models are compared to those effectively measured.

Erythrocytes, or red blood cells (RBCs), serve to deliver oxygen to cells. They take up oxygen in the lungs and release it while passing through the body's capillaries. Their cytoplasm is rich in hemoglobin, an iron-containing molecule that can bind oxygen and gives the characteristic red color to the blood. Mammalian erythrocytes are unique among the vertebrates as they are non-nucleated cells in their mature form. In the vast majority of mammalian species the typical shape is the biconcave disc, flattened and depressed in the center, a torus-shaped rim on the edge of the disc and with a dumbbell-shaped cross section (fig. 1). Its distinctive biconcave shape results in a high surface-area-to-volume (SA/V) ratio. This, together with deformability and mechanical stability of the membrane, allows the erythrocyte to undergo extensive passive deformation (Mohandas and Gallagher, 2008). Reduction of the SA/V ratio results in decreased cellular deformability, compromised red cell function, osmotic fragility and lessened survival. This is the case of hemolytic anemias, sickle cell disease, thalassemia, malaria, septicemia, diabetes mellitus (Mokken et al., 1992).

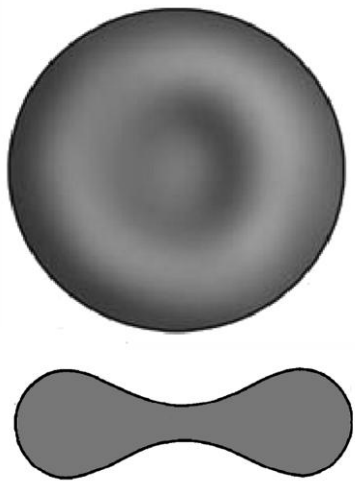

**Figure 1 – Erythrocyte surface and cross section**

Thus, knowledge of volume and surface area of erythrocytes is essential to evaluate their rheological properties. Calculation of the first is a simple method, part of the standard complete blood count. It is called Mean Corpuscular Volume (MCV) in hematology, expressed in femtoliters (fL=$10^{-15}$L=$\mu m^3$) and is calculated using the following formula:

$$MCV = 10 \times hematocrit/RBC\ count$$

The hematocrit is the volume percentage (%) of RBCs in blood and the RBC count is measured in millions/µL. Instead, calculation of the Surface Area (SA) is not so simple. It can be measured using special micropipettes (Waugh and Sarelius, 1996), but usually zoologists and veterinarians can rely only upon blood counts (automated or manual) and observations on blood smears.

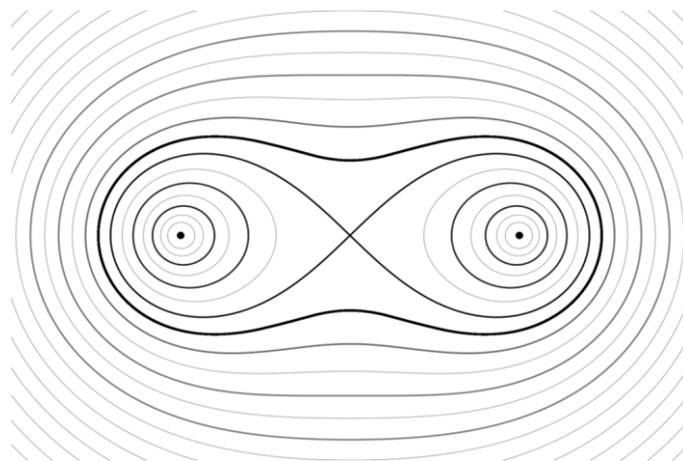

**Figure 2 – Cassini ovals**

The most precise description of the erythrocyte geometry is given by the equation of the oval of Cassini (fig. 2), following the physical explanation presented by Canham (1970). The main problem with the formulas proposed so far is the complexity of the mathematical expressions, which comprise elliptic integrals of the first and second kind (Vayo, 1983). For practical purpose it would be useful to find approximated models which give formulas based on diameter and thickness. It has been proposed to consider that the diameter squared of an erythrocyte determines the size of its surface area, while the thickness does not have a major influence (Kostelecka-Myrcha, 2002). This means considering the erythrocytes as a sphere (fig. 3b). Beside this model, we propose also a cylinder (fig. 3c) and an oblate spheroid with concave caps (fig. 3d).

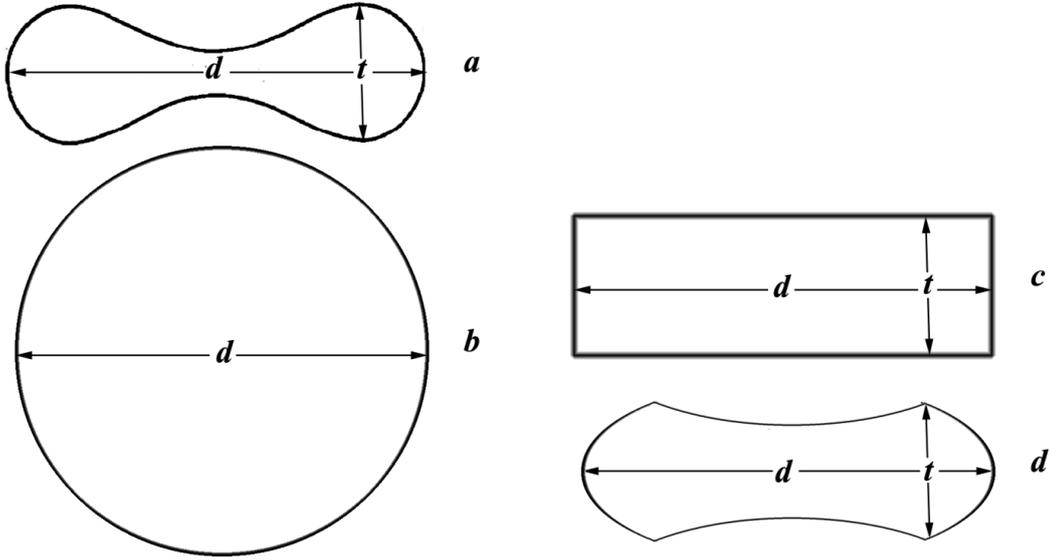

**Figure 3 – Cross section of erythrocyte (*a*) and proposed models: sphere (*b*), cylinder (*c*), spheroid with concave caps (*d*). *t* = thickness, *d* = diameter**

For the sphere, the formulas for volume (V) and Surface Area (SA) are:

$$V_s = \frac{\pi}{6} d^3$$

$$SA_s = \pi d^2$$

where *d* is the diameter and *t* the thickness. For the cylinder, the formulas are:

$$V_c = \frac{\pi}{4} d^2 t$$

$$SA_c = \pi d \left(\frac{d}{2} + t\right)$$

For the spheroid with concave caps, the formulas can be approximated as follows:

$$V_b = \frac{\pi}{4} t \left(d^2 - \frac{dt}{16} + \frac{t^2}{6}\right)$$

$$SA_b = \pi d \left(\frac{d}{2} + 2t \frac{\sinh^{-1} e}{e}\right)$$

where $e = \frac{2\sqrt{9d^2 - 4t^2}}{5t}$.

**Table 1 – Observed and calculated erythrocyte volumes in different species.**

| Species | d | t | MCV | $V_s$ | $V_c$ | $V_b$ | Reference |
|---|---|---|---|---|---|---|---|
| *Capra hircus* | 4.1 | 1.72 | 23.2 | 36.09 | 22.71 | 22.78 | Yamaguchi et al., 1987 |
| *Sorex araneus* | 4.51 | 1.62 | 24.5 | 48.03 | 25.88 | 25.85 | Wołk, 1974 |
| *Microtus arvalis* | 5.1 | 1.5 | 37.7 | 69.46 | 30.64 | 30.52 | Kostelecka-Myrcha, 1966 |
| *Microtus subterraneus* | 4.87 | 1.81 | 33.7 | 60.48 | 33.71 | 33.71 | Kostelecka-Myrcha, 1966 |
| *Lagurus lagurus* | 5.08 | 1.79 | 36 | 68.64 | 36.28 | 36.23 | Kostelecka-Myrcha, 1966 |
| *Myodes glareolus* | 5.1 | 1.85 | 37.6 | 69.46 | 37.79 | 37.76 | Kostelecka-Myrcha, 1967 |
| *Otomops martiensseni* | 5.3 | 1.8 | 40.7 | 77.95 | 39.71 | 39.63 | Kinoti, 1973 |
| *Microtus agrestis* | 5.04 | 2.07 | 41.1 | 67.03 | 41.29 | 41.39 | Kostelecka-Myrcha, 1966 |
| *Sicista betulina* | 5.6 | 1.7 | 42.8 | 91.95 | 41.87 | 41.72 | Wołk, 1985 |
| *Microtus oeconomus* | 5.08 | 2.16 | 44 | 68.64 | 43.78 | 43.93 | Wołk, 1970 |
| *Mus musculus* | 5.5 | 2.1 | 52.1 | 87.11 | 49.89 | 49.91 | Everds, 2006 |
| *Bison bonasus* | 5.77 | 2.1 | 54.84 | 100.58 | 54.91 | 54.87 | Wołk, 1983 |
| *Oryctolagus cuniculus* | 6.3 | 2.27 | 70.4 | 130.92 | 70.76 | 70.70 | Campbell, 2004 |
| *Homo sapiens* | 7.2 | 2.25 | 90 | 195.43 | 91.61 | 91.31 | Turgeon, 2004 |

**$d$=diameter, $t$=thickness, MCV=Mean Corpuscular Volume, $V_s$=Volume of the sphere, $V_c$=Volume of the cylinder, $V_b$=Volume of the biconcave spheroid.**

To test this models, we can use the few works where diameter, thickness and volume are directly measured. Table 1 compares the volumes obtained with the three models to the volumes measured as MCV in different mammalian species. The volume of the sphere is highly divergent from the measured volume. Instead, the volumes of the cylinder and the spheroid with concave caps are nearly identical and give a good approximation of the real volume, both with small and large erythrocytes.

**Table 2 – Surface areas of erythrocytes in different species**

| Species | d | t | $SA_m$ | $SA_s$ | $SA_c$ | $SA_b$ | Reference |
|---|---|---|---|---|---|---|---|
| *Capra hircus* | 4.1 | 1.72 | | 52.81 | 48.56 | 54.40 | Yamaguchi et al., 1987 |
| *Sorex araneus* | 4.51 | 1.62 | | 63.90 | 54.90 | 58.74 | Wołk 1974 |
| *Microtus arvalis* | 5.1 | 1.5 | | 81.71 | 64.89 | 66.02 | Kostelecka-Myrcha, 1966 |
| *Microtus subterraneus* | 4.87 | 1.81 | | 74.51 | 64.95 | 70.15 | Kostelecka-Myrcha, 1966 |
| *Lagurus lagurus* | 5.08 | 1.79 | | 81.07 | 69.10 | 73.54 | Kostelecka-Myrcha, 1966 |
| *Myodes glareolus* | 5.1 | 1.85 | | 81.71 | 70.50 | 75.63 | Kostelecka-Myrcha, 1967 |
| *Otomops martiensseni* | 5.3 | 1.8 | | 88.25 | 74.09 | 78.08 | Kinoti, 1973 |
| *Microtus agrestis* | 5.04 | 2.07 | | 79.80 | 72.68 | 80.88 | Kostelecka-Myrcha, 1966 |
| *Sicista betulina* | 5.6 | 1.7 | | 98.52 | 79.17 | 81.14 | Wołk, 1985 |
| *Microtus oeconomus* | 5.08 | 2.16 | | 81.07 | 75.01 | 84.38 | Wołk, 1970 |
| *Mus musculus* | 5.5 | 2.1 | 90.9 | 95.03 | 83.80 | 91.23 | Waugh and Sarelius, 1996 |
| *Bison bonasus* | 5.77 | 2.1 | | 104.59 | 90.36 | 97.03 | Wołk, 1983 |
| *Oryctolagus cuniculus* | 6.3 | 2.27 | | 124.69 | 107.27 | 114.86 | Campbell, 2004 |
| *Homo sapiens* | 7.2 | 2.25 | 136 | 162.86 | 132.32 | 136.55 | Turgeon, 2004 |

**$d$=diameter, $t$=thickness, $SA_m$=measured Surface Area, $SA_s$= Surface Area of the sphere, $SA_c$= Surface Area of the cylinder, $SA_b$= Surface Area of the biconcave spheroid.**

Unfortunately, the number of species used for direct measures of surface area are just two, man and mouse. Nonetheless, we present a comparison of these with the calculations obtained by the above formulas (tab.2). As can be seen, the sphere model gives an overestimation of the surface area of 5% in mouse and 20% in man. The cylinder model gives an underestimation of 8% in mouse and 3% in man. Finally, the biconcave spheroid gives an overestimation of 0.4% both in mouse and man.

From these data, it seems that thickness is a relevant measure in the determination not only of the volume, but also of the surface area. Therefore the spherical model should be rejected. While the volumes obtained with the cylinder and the biconcave spheroid models give a good approximation of the real volume and are nearly identical between them, the latter model seems more accurate in order to calculate the surface area.

As stated earlier, volume (MCV) and diameter are easily obtained from complete blood count and observations of blood smears, respectively. Thickness and surface area, instead, are much more difficult to be measured. Therefore, the following steps are proposed:

1. Measure of the diameter ($d$)
2. Determination of the MCV
3. Calculation of the thickness: $t = \frac{4MCV}{\pi d^2}$ (concerning the volume, the cylinder model is nearly identical to the spheroid, but is far more simple)
4. Calculation of the Surface Area

In this way, all the main geometric quantities of the erythrocyte are obtained and can be used in many fields such as comparative hematology, biorheology and veterinary medicine.

# REFERENCES

Campbell T.W. 2004. Mammalian hematology: Laboratory animals and miscellaneous species. In: Thrall M.A. (Ed.). Veterinary Hematology and Clinical Chemistry. Lippincott Williams and Wilkins, Philadelphia, PA, pp. 211-224.

Canham P.B. 1970. The minimum energy of bending as a possible explanation of the biconcave shape of the human red blood cell. J. Theor. Biol. 26(1): 61-81.

Everds N. 2006. Hematology of the laboratory mouse. in: Fox J., Barthold S., Davisson M., Newcomer C., Quimby F., Smith A. (Eds.). The Mouse in Biomedical Research, 2nd edition, Vol. III. Normative Biology, Husbandry and Models. Elsevier Academic Press, San Diego, CA, pp. 133-170.

Kinoti G.K. 1973. Observations on the blood of a tropical bat, Otomops martiensseni. East African Wildlife Journal 11(2): 129-134.

Kostelecka-Myrcha A. 1966. Hemoglobin, erythrocytes and hematocrit in the blood of some microtidae under laboratory conditions. Bull. Acad. Pol. Sci. Biol. 14(5): 343-349.

Kostelecka-Myrcha A. 1967. Variation of morpho-physiological indices of blood in Clethrionomys glareolus (Schreber, 1780). Acta Theriol. 12: 191-222.

Kostelecka-Myrcha A. 2002. The ratio of amount of haemoglobin to total surface area of erythrocytes in mammals. Acta Theriol. 47(1): 209-220.

Mohandas N., Gallagher P.G. 2008. Red cell membrane: past, present, and future. Blood 112: 3939.

Mokken F.C., Kedaria M., Henny C.P., Hardeman M.R., Gelb A.W. 1992. The clinical importance of erythrocyte deformability, a hemorrheological parameter. Ann. Hematol. 64(3): 113-122.

Vayo H.W. 1983. Some red blood cell geometry. Can. J. Physiol. Pharmacol. 61(6): 646-649.

Waugh R.E., Sarelius I.H. 1996. Effects of lost surface area on red blood cells and red blood cell survival in mice. Am. J. Physiol. Cell Physiol. 271: C1847-C1852.

Wołk E. 1970. Erythrocytes, haemoglobin and haematocrit in the postnatal development of the root vole. Acta Theriol. 15: 283-293.

Wołk E. 1974. Variations in the hematological parameters of shrews. Acta Theriol. 19: 315-346.

Wołk E. 1983. The hematology of the free-ranging European bison. Acta Theriol. 28: 73-82.

Wołk E. 1985. Hematology of a hibernating rodent - the northern birch mouse. Acta Theriol. 30: 337-348.

Yamaguchi K., Jürgens K.D., Bartels H., Piiper J. 1987. Oxygen transfer properties and dimensions of red blood cells in high-altitude camelids, dromedary camel and goat. J. Comp. Physiol. B 157(1): 1-9.